\documentclass[lettersize,journal]{IEEEtran}
\usepackage{amsmath,amsfonts}
\usepackage{algorithmic}
\usepackage{algorithm}
\usepackage{array}
\usepackage[caption=false,font=normalsize,labelfont=sf,textfont=sf]{subfig}
\usepackage{textcomp}
\usepackage{stfloats}
\usepackage{url}
\usepackage{verbatim}
\usepackage{graphicx}
\usepackage{cite}
\usepackage{multirow}
\usepackage{makecell}
\usepackage{pifont}
\usepackage{flushend} 
\begin{document}

\title{Intelligent Distributed Optical Fiber Sensing in Transportation Infrastructures: Research Progress, Applications, and Challenges}

\author{Xin Gui, Fanhao Zeng, Yunchuan Zhang, \IEEEmembership{Member,~IEEE}, Yiming Wang, Jiaqi Wang, Changjia Wang, Xuelei Fu, Sheng Li, Fang Liu, Lina Yue, Jinpeng Jiang and Zhengying Li
\thanks{This work was supported by the National Natural Science Foundation of China under Grant No. 62471347, and the Open Fund of Hubei Longzhong Laboratory under 2024KF-01.\textit{(Corresponding author: Zhengying Li.)}}
\thanks{Xin Gui and Jiaqi Wang are with Hubei Longzhong Laboratory, Wuhan University of Technology Xiangyang Demonstration Zone, Xiangyang 441000, China; National Engineering Research Center of Fiber Optic Sensing Technology and Networks, Wuhan University of Technology, Wuhan 430070, China(email: guixin@whut.edu.cn; wjq@whut.edu.cn);\\
Sheng Li, Fang Liu, Lina Yue and Jinpeng Jiang are with National Engineering Research Center of Fiber Optic Sensing Technology and Networks, Wuhan University of Technology, Wuhan 430070, China(email: lisheng@whut.edu.cn; fangliu@whut.edu.cn; linayue@whut.edu.cn; jiangjp2812@whut.edu.cn);\\
Fanhao Zeng, Yunchuan Zhang, Yiming Wang, Changjia Wang, Xuelei Fu and Zhengying Li are with School of Information Engineering, Hubei Key Laboratory of Broadband Wireless Communication and Sensor Networks, Wuhan University of Technology, Wuhan 430070, China (email: fhzeng@whut.edu.cn; yunchuan.zhang@whut.edu.cn; wangyiming@whut.edu.cn; wangchangjia@whut.edu.cn; xlfu@whut.edu.cn; zhyli@whut.edu.cn).}}



\maketitle

\begin{abstract}
Distributed optical fiber sensing (DOFS), along with its capabilities of long-range coverage, multi-parameter monitoring, and completely passive detection, emerges as one of the most promising non-destructive detection techniques for structural health monitoring (SHM) and operational assessment of linear transportation infrastructures. In this paper, we provide a state-of-the-art review on DOFS applications across typical linear infrastructure systems, encompassing highways, long-span bridges, rail transit networks, airport runways, and analogous linear structures. The comprehensive discussion consists of four critical research dimensions: 1) optical fiber selection for multi-parameter sensing and robust cable packaging techniques, 2) distributed sensing principles and signal processing algorithms, 3) diverse application scenarios in SHM and related fields, and 4) anomaly detection and event classification methodologies. Building upon the foundational introduction of DOFS technical principles and monitoring solutions for intelligent transportation infrastructure, this paper elaborates on system design approaches, sensing data analytics algorithms, and future research directions. 
\end{abstract}


\section{Introduction}
\IEEEPARstart{T}{he} health and operational status monitoring of large-scale transportation infrastructures holds significant importance for ensuring stable economic and social operation, and has become a global research focus drawing extensive social attention. Taking China as an example, by the end of 2024, the total mileage of the national comprehensive transportation network has exceeded 6 million kilometers, including over 5.4 million kilometers of highways, more than 162,000 kilometers of railways, and 1.1 million bridges, forming a comprehensive transportation system centered on high-speed networks\cite{101Transportation}. However, during long-term service, these infrastructures are subjected to coupled effects from complex geological environments, extreme climates, and high-frequency loads, leading to increasingly prominent issues such as subgrade voids, bridge-tunnel settlement deformation, slope instability, which severely threaten transportation safety.

Conventional monitoring techniques (such as manual inspections and electrical sensor-based monitoring) face limitations in monitoring range, cumbersome cabling, and susceptibility to electromagnetic interference, rendering them inadequate to meet the requirements of long distance, all-round coverage, low latency, and continuously reliable monitoring against environmental changes \cite{102khan2016integration}. In contrast, distributed optical fiber sensing (DOFS) technology has emerged as one of the most promising online non-destructive detection methods for monitoring the health of long-linear transportation infrastructures due to its advantages on long distance capability, high spatial resolution, electromagnetic interference resistance, and passive operation without power supply. By using optical fibers as transmission media and sensing elements simultaneously, DOFS captures variations in signal intensity, phase, wavelength, or polarization. This capability enables the distributed monitoring of temperature, strain, and acoustics, providing a comprehensive assessment of the structural integrity of large-scale linear infrastructures. \cite{103bado2021review,104gui2023distributed}. 

Recent advances show that DOFS-aided transportation infrastructures monitoring can be extended to diverse deployments and application scenarios. \cite{105wijaya2021distributed,106wang2023applications,107zhang2024distributed}. Nevertheless, significant technical challenges remain in practical deployments, mainly involving optimized fabrication of specialty optical fibers, engineering-adapted optical cable designs, massive monitoring data processing, and reliability validation for long-term service performance \cite{108westbrook2020big}. On the other hand, the large-scale spatiotemporal sensing data acquired by DOFS systems provides crucial foundations for artificial intelligence (AI) and machine learning (ML) applications in infrastructure monitoring \cite{109ohodnicki2022fusion,110liu2023intelligent}. The parallel advancement of these technologies enables breakthroughs in intelligent processing of massive sensing data and complex pattern recognition, allowing for predictive assessments prior to structural deterioration which significantly enhances the overall efficiency and reliability of monitoring systems.

To illustrate the solutions to the aforementioned challenges, this paper systematically reviews recent advances in DOFS technology and its typical applications in transportation infrastructures monitoring. To start, we thoroughly analyze the key theoretical issues including DOFS operating principles, system architectures, and domain-specific technical requirements for transportation. Next, we present the technical breakthroughs such as intelligent sensing frameworks that leverage ML algorithms for processing massive sensor datasets. By integrating representative engineering case studies across highways, urban rail systems, and bridges, this paper further explores future technical trends. It aims to provide systematic references for both theoretical research and engineering applications in intelligent monitoring of long-linear transportation infrastructures.

\section{Sensing Optical Fiber Preparation and Sensing Optical Cable Packaging Methods}\label{2}
DOFS is widely applied to structural health and operational monitoring of large-scale infrastructures, owing to its advantages of long distance monitoring capabilities. However, it still encounters two major challenges. First, while traditional communication optical fibers exhibit scattering effects, the backscattered signals are inherently weak and attenuated significantly as transmission distance increases. This results in a low signal-to-noise ratio (SNR) for sensing signals in long distance distributed sensing systems, thereby containing the corresponding remote sensing performance. In this context, the primary technical hurdle lies in boosting scattering signals while maintaining low transmission loss, alongside the structural reconfiguration of communication fibers for specialized sensing applications. Second, in complex engineering environments, sensing optical fibers must be encapsulated in cables to ensure mechanical strength and environmental adaptability. But this encapsulation may introduce issues such as strain transfer loss and temperature response hysteresis. Striking the balance between engineering feasibility and sensing accuracy remains a key challenge in optical fiber packaging design.

In this chapter, we investigate the current advances in sensing optical fiber design and fabrication, including scattering-enhanced fibers, Fiber Bragg Grating (FBG) array sensing fibers, and micro-structured sensing fibers. In addition, we provide a detailed overview of optical cable packaging technologies, such as efficient strain transfer structures, temperature compensation, and techniques to improve sensing sensitivity to acoustic pressure.

\subsection{DOFS Based on Fiber Bragg Grating Array}\label{2.1}
Scattering-enhanced specialty optical fibers are introduced into distributed sensing systems to overcome the limitations of conventional single-mode communication fibers in distributed sensing, particularly in terms of specific sensing parameters and overall performance. By continuously modifying the fiber materials, structures, or incorporating discrete scattering enhancement mechanisms, these fibers achieve true sensing functionality, enabling significant improvements in sensitivity and precision.
\subsubsection{Scattering-Enhanced Sensing Fibers for Increasing the Scattering Coefficient}\label{2.1.1}
Boosting the scattering effects within the optical fiber to improve the SNR is a common method. The most straightforward way to increase the backscattering in optical fibers is to introduce more inhomogeneities, thereby generating additional scattered light. Typical approaches include fiber irradiation, doping with nanoparticles, and employing special fiber structures.

\textbf{Irradiation Modification Technology}: This scheme uses the photosensitivity of germanium-doped silica fibers, where ultraviolet (UV) light exposure modifies the defect structures within the fiber, thereby inducing refractive index modulation and enhancing scattering. However, this continuous exposure approach encounters several challenges, including high fabrication costs, low production speed (on the order of hundreds of micrometers per second), and the requirements to remove the fiber coating, which restrict its engineering feasibility. Experimental results have shown that this technique can increase scattering intensity by one to two orders of magnitude\cite{3106loranger2016enhancement} and improve the system SNR by more than 20 dB\cite{3107redding2020low,3108loranger2015rayleigh,3109du2023high}. In 2020, OFS (USA) demonstrated a low-loss scattering-enhanced optical fiber with a length of 1.5 km and an attenuation level of 0.5 dB/km\cite{3110westbrook2020enhanced}, effectively addressing the problem of excessive fiber attenuation. 

\textbf{Nanoparticle Doping Technology}: Nanoparticle doping technology significantly improves the SNR of backscattered light by incorporating oxides such as boron and germanium during the optical fiber fabrication process\cite{3111tsai1989correlation,3112sypabekova2018fiber}. A key advantage of this approach is the compatibility with the standard fiber drawing process which preserves the mechanical properties of the sensing optical fiber. However, its application in long distance sensing is still constrained by high transmission losses. For instance, magnesium oxide-doped fibers are relatively simple to manufacture and can be designed to match the dimensions of standard single-mode fibers (SMF), but transmission loss remains a significant challenge, with attenuation reaching up to 14.3 dB/m\cite{3113tosi2021rayleigh}. Research led by the Canadian team under Fuertes\cite{3114fuertes2021engineering,3115fuertes2023tunable,3116fuertes2023customizing} demonstrated that by optimizing the size, morphology, and distribution of Ga and Ba-doped nanoparticles during the preform and fiber fabrication processes, fiber transmission loss could be reduced to 0.1–0.2 dB/m. These findings extend the sensing range to hundreds of meters, with scattering enhancement and detection performance that exceed those of Mg-doped optical fibers. 

\textbf{Numerical Aperture (NA) Optimization Technology}: This line of works aim to improve backscattering performance by increasing the NA of optical fibers. Specifically, polymer optical fibers with larger scattering cross-section can achieve over a 30 dB higher scattering enhancement compared to conventional silica fibers\cite{3117dengler2021absolute}. Due to material limitations, their attenuation is still excessively high, ranging from 9.2 to 15 dB/km, which limits their applications to shorter sensing distances\cite{3118sugita2001optical,3119lenke2008improvements}. Multimode optical fibers (MMF)\cite{3120zhang2012optimized,3121ekechukwu2023degradation,3122guo2012novel,3123li2022physics} achieve a 2–3-fold increase in backscattering intensity by raising the germanium doping concentration in the core. The typical transmission loss for these fibers ranges from 0.29 to 0.35 dB/km. Prior arts on multicore optical fibers\cite{3124westbrook2014integrated} have demonstrated that increasing the core’s numerical aperture from 0.13 to 0.21 results in a 3 dB gain in backscattering performance. However, this comes with the price of significantly higher fiber transmission loss, exceeding 1 dB/km.

\begin{table*}[htbp]
	\centering
	\caption{Current Development Status of Scattering Enhanced Fiber\\by Increasing Scattering Coefficient}\label{tab3}
	\renewcommand\arraystretch{2} 
	\begin{tabular}{ccccccc}
		\hline
		\textbf{Category} & \textbf{Method} & \textbf{Year} & \textbf{\makecell[c]{Scattering\\enhancement(dB)}} & \textbf{Loss} & \textbf{Length} & \textbf{Ref.} \\
		\hline
		\multirow{4}*{\textbf{Irradiation}}&\multirow{4}*{UV}& 2015 & 25.0 & -- & -- & \cite{3108loranger2015rayleigh} \\
		\cline{3-7}
		 &   & 2016 & 17.0--20.0 & --         & 8.0\,m   & \cite{3106loranger2016enhancement} \\
		\cline{3-7}
		 &   & 2020 & 7.0        & 0.5\,dB/km & 40.0\,km & \cite{3110westbrook2020enhanced} \\
		\cline{3-7}
	     &   & 2023 & 37.3       &     --     & 10.0\,m  & \cite{3109du2023high} \\
		\hline
		\multirow{4}*{\textbf{Doping}}& MgO & 2021 & 48.9 & 14.3\,dB/m  & 3.0\,m & \cite{3113tosi2021rayleigh} \\
		\cline{2-7}
		 & \multirow{2}*{Ca} & 2021 & 25.9--44.9 & 0.1--8.7\,dB/m & 32.1\,m & \cite{3114fuertes2021engineering} \\
		\cline{3-7}
		 &       & 2023 & 26.4--43.8 & 0.2--5.1\,dB/m & 57.5\,m & \cite{3115fuertes2023tunable} \\
		\cline{2-7}
		 &  Ba   & 2023 & 26.4--43.8 & 0.2--5.1\,dB/m & 57.5\,m & \cite{3116fuertes2023customizing} \\
		\hline
		\multirow{4}*{\textbf{NA}}& \makecell{Polymer\\optical fibers} & 2017 & 30.0 & 9.2--11.3\,dB/m  & 5.0--20.0\,m & \cite{3117dengler2021absolute} \\
		\cline{2-7}	
		 & \multirow{2}*{MMF} & 2012 & 1.7--3.8 & -- & 10.0\,km & \cite{3120zhang2012optimized} \\
		\cline{3-7}
		 &   & 2012 & 2.0--4.8 & 0.29-0.35\,dB/km & --   & \cite{3122guo2012novel} \\
		\cline{2-7}
		 & \makecell{Multicore\\fibers} & 2014 & 3.0 & 1.0\,dB/km  & 10.0\,m  & \cite{3124westbrook2014integrated} \\
		\hline
	\end{tabular}
\end{table*}

\subsubsection{Sensing Fibers Based on Laser Modification Technology}\label{2.1.2}
Laser modification technologies in the field of Optical Fiber Sensing (OFS) mainly include two approaches: UV laser irradiation and femtosecond laser inscription. UV laser irradiation utilizes the photosensitivity of fiber materials to induce refractive index modulation through UV exposure, while femtosecond laser inscription achieves permanent refractive index modification via the nonlinear absorption effect of high peak-power laser pulses. Compared with UV-based methods, femtosecond inscription offers advantages such as no need to remove the coating layer and excellent high-temperature resistance. In the fabrication of scattering-enhanced fibers, UV laser irradiation remains the dominant technique, primarily enhancing performance through the inscription of ultra-weak FBG arrays or colorless weak-reflection point arrays.

\begin{figure*}[htbp]
	\centering
    \includegraphics[width=7in]{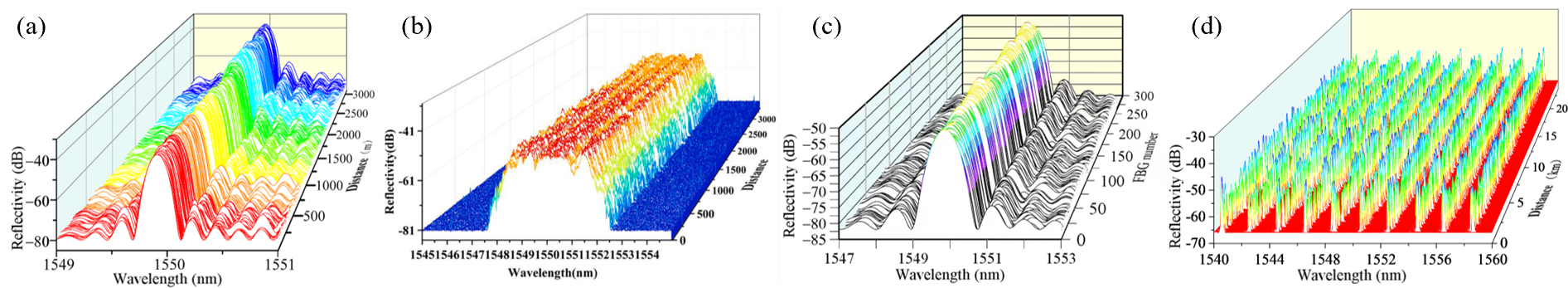}
	\caption{Spectra of various types of UWFBG array fabricated via the drawing tower process. (a) Apodized FBGs array, (b) chirped FBGs array, (c) ultra-short FBGs array, (d) multi-wavelength FBGs array.}
	\label{fig1}
\end{figure*}

In the field of FBG array fabrication, research efforts mainly focus on 3 approaches: online fabrication on fiber drawing towers, online fabrication using UV-transparent coated fibers, and femtosecond laser inscription. Among them, the online fabrication on drawing tower method is currently the most mature industrial solution, which adopts two main techniques: the UV phase mask method and the UV Talbot interferometer method. This method originates from a 1993 study at the University of Southampton, UK, which demonstrated the use of a KrF excimer laser to inscribe gratings in situ during the fiber drawing process\cite{3125dong1993single}. Their proposed approach not only preserves the mechanical strength of the fiber but also simplifies the conventional FBG fabrication process. In a collaboration between our research group and FBGs Company (Germany), the commercial production of weak FBG arrays has been achieved. A fabrication process and supporting equipment based on the phase mask technique have been developed, enabling the integration of up to 100,000 ultra-weak FBGs (UWFBG) on a single fiber, with a wavelength consistency deviation within ±20 pm\cite{104gui2023distributed}. With innovations such as optimized fiber compositions and high-temperature-resistant coatings, a sensing optical fiber of long-term stability at 350 °C has been developed. Breakthroughs have also been made in fine-diameter (80 $\mu$m) fiber fabrication, low bending loss ($\leq$ 0.05 dB after 25 turns at a 10 mm diameter for a 1550 nm wavelength), and multi-wavelength (10-wavelength) switching. Furthermore, stable fabrication of various FBG, including apodized and chirped FBGs, has been realized, enabling successful application in multi-parameter long-distance sensing, as shown in Fig. \ref{fig1}.

The UV-transparent coated optical fiber has achieved remarkable progress in recent years. In 2017, OFS (USA) developed an innovative phase mask-based approach, which combined with a high-precision positioning control system, enabled kilometer-scale array fabrication\cite{3127westbrook2017kilometer}. In 2021, Xiao X P team at Huazhong University of Science and Technology developed a high-performance coating with 79\% UV transmittance, which significantly enhanced fabrication precision and parameter control capability\cite{3128xiao2022line} when integrated with the phase mask technique. 

In addition to enabling customized fabrication of conventional FBG, femtosecond laser technology demonstrates unique advantages in special application scenarios such as high-temperature sensing. Compared with the 450 °C temperature limit of traditional phase mask-based drawing tower techniques, femtosecond laser technology lifts the material constraints and allows high-quality grating inscription on non-silica substrates such as sapphire, achieving temperature resistance exceeding 1000 °C. Professor Wang Yiping’s team at Shenzhen University has made significant breakthroughs in this field\cite{3129meng2022submillimeter}. They developed a novel fabrication system based on femtosecond laser point-by-point (PbP) technology which enables the production of high temperature resistant, weak reflectivity, and FBG arrays with tunable wavelength. The proposed system succeeds in solving the key technical problems of conventional femtosecond laser inscription, which previously struggled to balance large scale multiplexing requirements and DAS applications with limited spectrum bandwidth.

In terms of reflective point array technology, both UV exposure and femtosecond laser inscription techniques can be employed to fabricate reflection point arrays. Wang Yiping’s team achieved a 26 dB enhancement in Rayleigh backscattering in single-mode fibers using femtosecond laser processing\cite{3129meng2022submillimeter}. Sun Qizhen’s team at Huazhong University of Science and Technology fabricated micro-structured fibers with 5 m spacing and a 15 dB enhancement in backscattering using UV exposure technology\cite{3130liu2020new}. Zhang Jianzhong’s team at Harbin Engineering University realized a 30 dB SNR enhancement and achieved temperature sensing up to 1000 °C through reflection points inscribed by femtosecond laser\cite{3131li2022fiber}. 

The technical advances and applications of these novel methods have made significant strides in addressing fundamental challenges such as transmission loss and high temperature tolerance. These breakthroughs are propelling DOFS techniques towards an expanded operational temperature range from cryogenic to ultra-high temperatures, longer sensing distances, higher spatial resolution, and improved environmental adaptability. The synergistic developments of diverse fabrication techniques, along with their novel applications, are establishing a robust technical foundation for the large scale engineering implementations of OFS networks.

\subsection{Distributed Sensing Optical Cables}\label{2.2}
When employing fiber-optic distributed sensing technology for fully distributed SHM and conditional monitoring of large scale transportation infrastructures, it is necessary to protect the sensing fibers through encapsulation to withstand harsh outdoor environments and challenging construction conditions. Optical cables used in such scenarios primarily target three detection parameters: temperature measurement, strain measurement, and vibration measurement. The sensing methods are mainly categorized into fiber scattering-based distributed detection and fiber grating array-based quasi-distributed detection, each employing distinct encapsulation approaches. In this section, we investigate these two sensing methods in relation to the three types of physical quantities: temperature, strain, and vibration.

\subsubsection{Distributed Temperature Sensing Optical Cables}\label{2.2.1}
DOFS approaches build upon the principle of backscattering and employ technical frameworks such as time-division multiplexing (TDM), wavelength-division multiplexing (WDM), and frequency-division multiplexing (FDM). In general, the sensing media consists of types: conventional single-mode fibers, scattering-enhanced fibers, and FBG arrays. Raman scattering technology is only sensitive to temperature, and its encapsulation design emphasizes temperature response characteristics. In contrast, Rayleigh/Brillouin scattering and FBG technologies exhibit temperature-strain cross-sensitivity. Stress-free solutions, such as internal loose tubes\cite{3201li2021deformation} or adhesive encapsulation\cite{3202wang2020high}, are typically adopted to eliminate strain interference.

Regarding engineering applications, surface-mounted designs are mainly used for fire monitoring in enclosed spaces such as tunnels\cite{3203sun2022fiber} and urban rail transit systems. These designs employ flat or semi-circular thermally conductive sheath structures, emphasizing construction convenience and environmental durability. Embedded designs are applied for monitoring within concrete structures\cite{3204zhou2018feedback}, where armored reinforcements ensure long-term reliable operations. Scattering-based techniques excel in monitoring large scale temperature fields, while FBG array approaches hold advantages in refined monitoring scenarios, e.g., local hot spot detection. Beyond transportation infrastructures, these techniques also have obtained superior performance in long distance pipelines, power facilities, and building structures.

Practical deployments still necessitate further optimization of the compatibility of sensing solutions. The sensing characteristics and structures of temperature sensing cables are shown in Table \ref{tab4}.

\begin{table*}[htbp]
	\caption{The Sensing Characteristics and Structures of Temperature Sensing Cables.\label{tab4}}
	\centering
	\begin{tabular}{cclcc}
		\hline
		\textbf{Sensing Principle} & \textbf{Cable Type} & \textbf{Typical Structure} & \textbf{Manufacturing} & \textbf{\makecell{Measurement\\Time}}  \\
		\hline
		Raman Scattering & \makecell{High Mechanical\\Strength Cable\cite{3204zhou2018feedback}} & 
		\begin{minipage}[b]{0.5\columnwidth}
			\centering
			\raisebox{-.5\height}{\includegraphics[width=\linewidth]{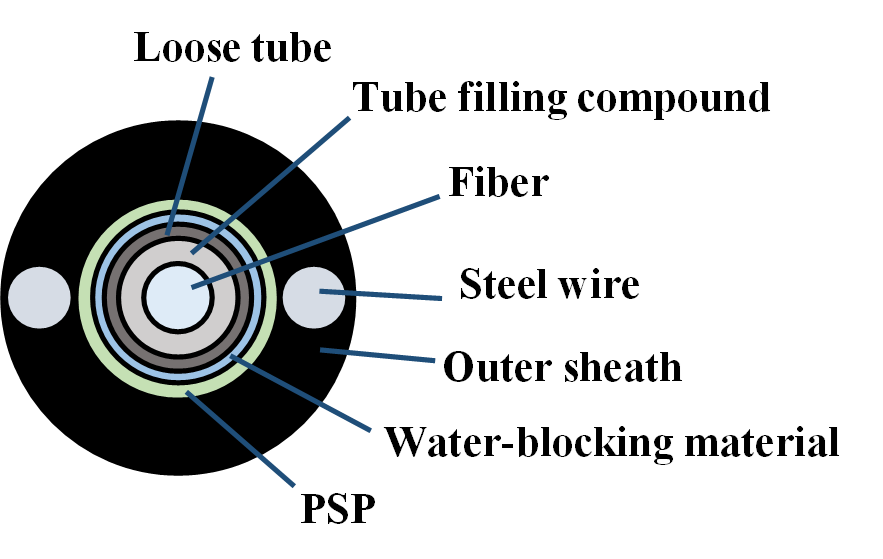}}
		\end{minipage}
		& \makecell{Water-blocking Filling
		\\Steel Wire Strength Member} & $\approx$ 60\,s\\
		\hline
		\makecell{\\ \\ \\ Brillouin Scattering \\ \\ \\} & \multirow{3}*{\makecell{\\ \\ \\ \\ \\ \\Strain-Free Cable\cite{3201li2021deformation,3202wang2020high}}} 
		& \multirow{2}*{
		\begin{minipage}[!t]{0.4\columnwidth}
			\centering
			\raisebox{-.5\height}{\includegraphics[width=\linewidth]{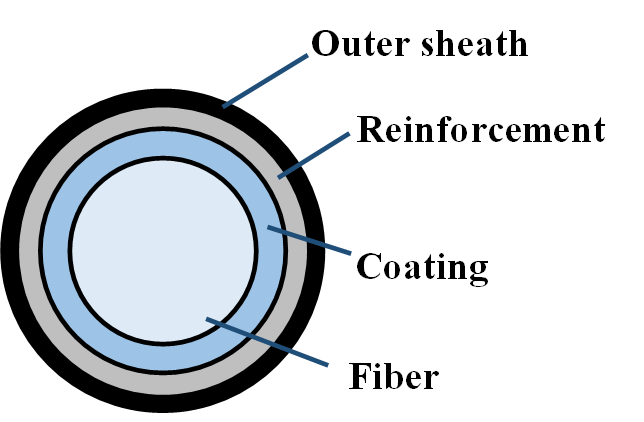}}
		\end{minipage}}
		 &\multirow{2}*{\makecell{\\ \\ \\Loose-Tube Sheathing}} & 1--60\,s \\
		\cline{1-1}
		\cline{5-5}
		\makecell{\\ \\ \\ Rayleigh Scattering \\ \\ \\} & & & & 3\,min \\
		\cline{1-1}
		\cline{3-5}
		FBG Array& &
			\begin{minipage}[!t]{0.5\columnwidth}
			\centering
			\raisebox{-.5\height}{\includegraphics[width=\linewidth]{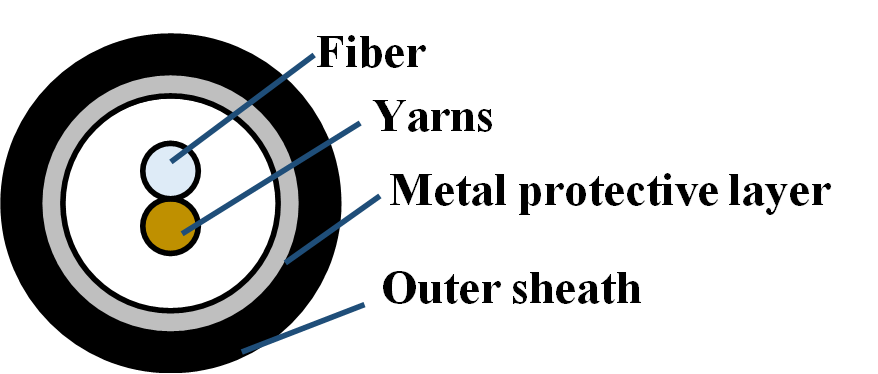}}
		\end{minipage}
		 & Internal Dispensing & $\le$ 1\,s \\
		\hline
	\end{tabular}
\end{table*}

\subsubsection{Distributed Strain Sensing Optical Cables}\label{2.2.2}
The encapsulation of strain sensing optical cables brings a major technical challenge in the field of OFS, i.e., realizing a probe-free encapsulation solution for long distance distributed strain detection. The key challenges can be categorized into the following aspects: 1) ensuring effective strain transfer over long distances and maintaining consistency across multiple distributed zones; 2) eliminating the interference of temperature variations on measurement results. Existing studies \cite{3205zheng2021strain,3206tan2021strain} indicate that, given the shear lag effect, the strain actually detected by the optical fiber is often lower than the true deformation of the measured object. This makes the material properties of the cable sheath, its geometric structure, and the performance of the bonding layer key factors affect strain transfer efficiency.

\begin{figure}[htbp]
	\centering
	\includegraphics[width=3.5in]{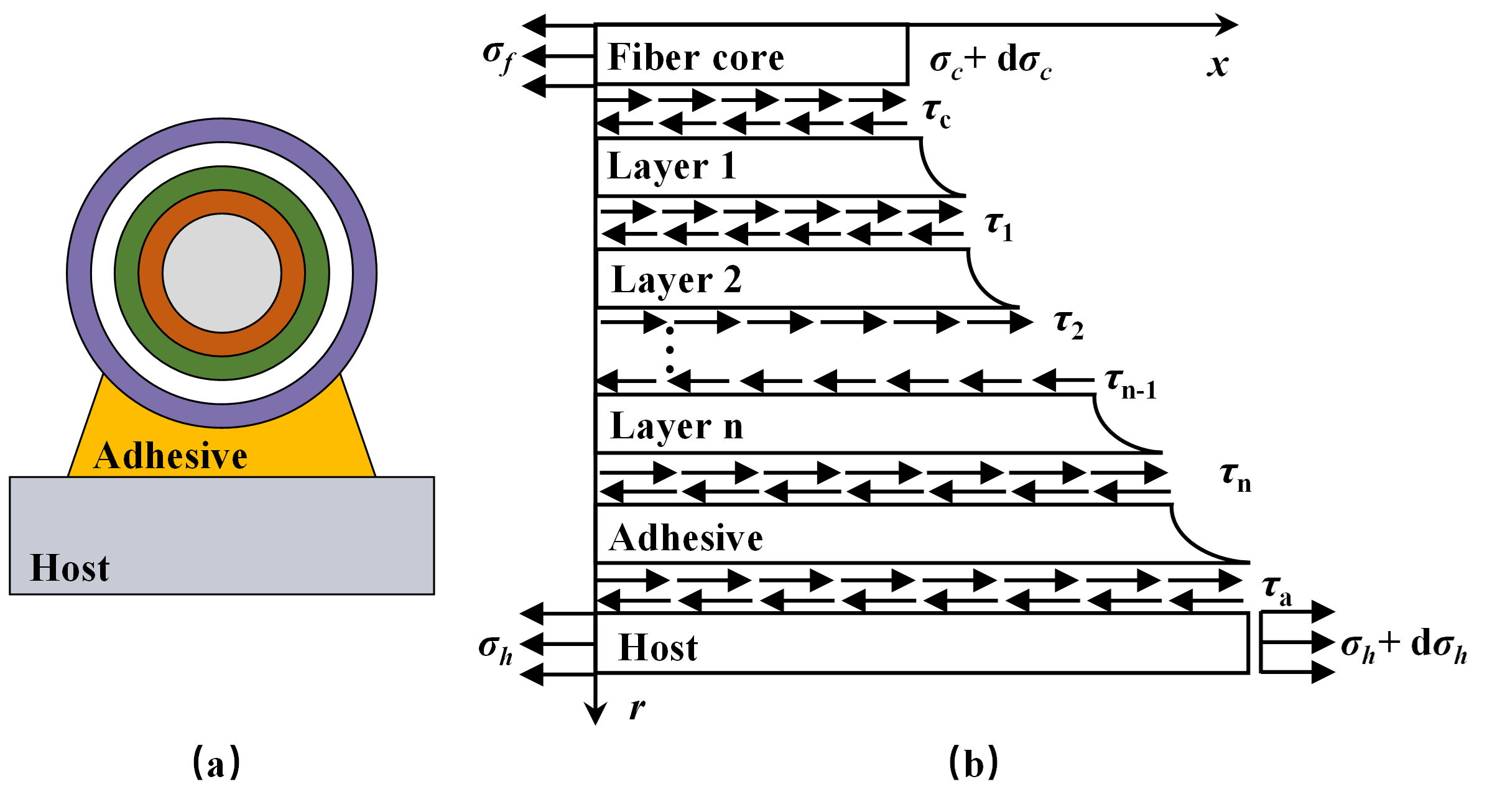}
	\caption{Strain transfer mechanism in a surface-bonded multi-layered distributed FO sensor. (a) Cross section. (b) Stress state of a cable element\cite{3205zheng2021strain}.}
	\label{fig2}
\end{figure}

To improve strain transfer efficiency, prior works employ a single-layer tight-buffered structure design and optimize encapsulation parameters through mechanical modeling. For scattering-based OFS, the axial non-uniformity of the strain transfer coefficient can be mitigated by global calibration or pre-embedded gauge length units\cite{3207chengcheng2019theoretical}. Regarding FBG array sensing, the long-gauge continuous encapsulation technology developed in \cite{3208yue2024research,3209nan2024experimental} overcomes the limitation of conventional FBG's discrete point measurement, achieving a strain monitoring solution that combines high SNR with distributed detection capability. Its flexible gauge length feature allows adaptation to the spatial resolution requirements of different application scenarios. For the purpose of addressing the issue of temperature interference, previous studies mainly involve dual-cable parallel deployments or integrated optical fiber schemes\cite{3201li2021deformation,3210falcetelli2020strain,3211zhang2020situ}. The former one provides a compensation reference through a dedicated temperature-sensing cable, while the latter utilizes composite cables to achieve simultaneous decoupling of temperature and strain. The sensing characteristics and structure of the strain-sensing cable are shown in Table \ref{tab5}.

The encapsulation technologies for distributed fiber optic strain sensing cables are often divided into surface-mounted and embedded methods, each with distinct characteristics in application scenarios and technical challenges. The key challenge for surface-mounted encapsulation lies in improving the strain transfer efficiency of the cable. Recent works address this issue by optimizing theoretical models and interface treatment methods. Zheng et al.\cite{3205zheng2021strain}, by establishing a strain transfer mechanism under strain gradient effects, confirmed that using coating materials with a high shear modulus can significantly enhance strain transfer efficiency. Liu et al.\cite{3212liu2022strain} proposed a finite element model-updated shear lag parameter correction method, effectively improving the accuracy of measurement data from surface-mounted sensing cables. In practical applications, the technology of embedding polyimide optical fibers into thermoplastic tape\cite{3213enckell2011evaluation} improves the coordinated deformation capability between the cable and the substrate. Woven fabric-embedded optical fiber encapsulation enhances frictional effects\cite{3214biondi2021optical,3215biondi2023smart}, and an embossed sheath structure overcomes the issue of easy detachment common in traditional adhesive methods through mechanical interlocking, thereby improving long-term stability\cite{3216alj2021environmental}.

Embedded encapsulation technology must simultaneously address two key performance metrics: strain transfer efficiency and mechanical strength. Research by Tan et al.\cite{3206tan2021strain} indicates that the strain transfer efficiency of embedded cables depends on parameters such as the elastic modulus of the host material, the adhesive, and the optical fiber itself. The three-layer strain transfer model incorporating viscoelastic behavior, established by Wang et al.\cite{3217wang2016strain}, provides theoretical guidance for reducing measurement errors in embedded applications. In practical application, the double-layer armored structure design proposed by our research team\cite{3218nan2025study} offers an effective solution for balancing mechanical protection and sensitivity in embedded sensing cables. This design features an inner layer comprising a stainless-steel spiral tube to protect the optical fiber, with parallel steel wires embedded alongside it. The outer layer employs a longitudinally wrapped steel tape to enhance compressive strength. A structural adhesive fills the intermediate space to ensure uniform stress transfer. The reliability of this technical approach was further validated by Li et al.\cite{3201li2021deformation} using a double-layer tight-buffered optical cable for tunnel lining monitoring. Their work demonstrated that well-optimized embedded sensing cables can accurately reflect the strain behavior of structures.

\begin{table*}[htbp]
	\caption{The Sensing Characteristics and Structures of Strain Sensing Cables.\label{tab5}}
	\centering
	\begin{tabular}{ccclc}
		\hline
		\textbf{Sensing Principle} & \textbf{Type} & \textbf{Cable Type} & \textbf{Typical Structure} & \textbf{\makecell{Installation\\Method}}  \\
		\hline
		Brillouin Scattering & Distributed & \makecell{Tight-Buffered\\Cable\cite{3201li2021deformation,3218nan2025study,3219yoon2016real}} &
		\begin{minipage}[t]{0.45\columnwidth}
			\centering
			\raisebox{-.5\height}{\includegraphics[width=\linewidth]{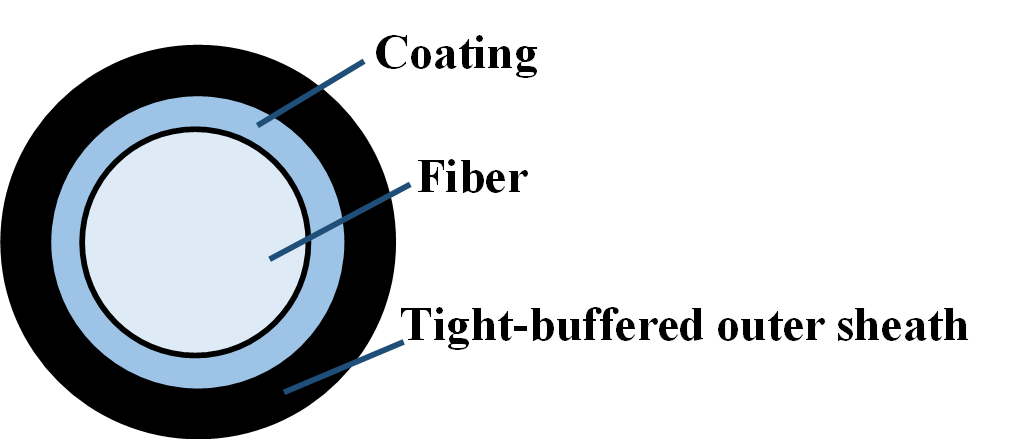}}
		\end{minipage}
		& Surface Embedded \\
		\hline
		Brillouin Scattering & Distributed & SMARTape\cite{3213enckell2011evaluation}
		&
		\begin{minipage}[t]{0.5\columnwidth}
			\centering
			\raisebox{-.5\height}{\includegraphics[width=\linewidth]{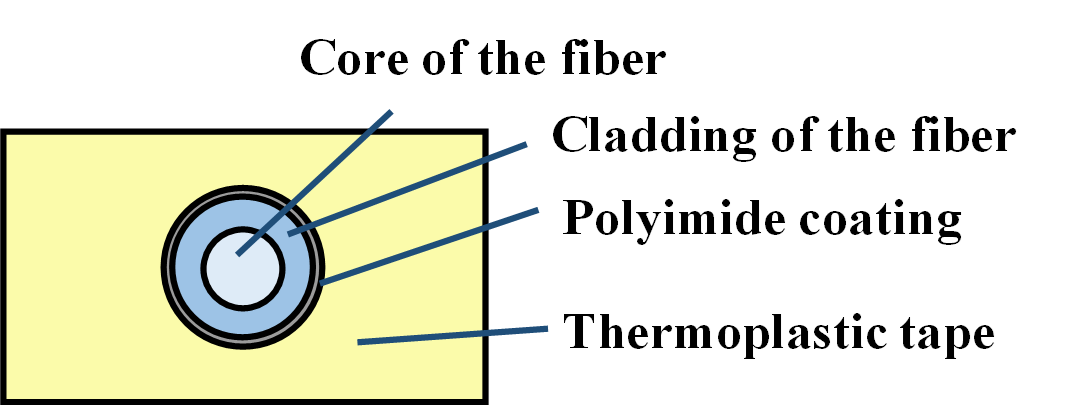}}
		\end{minipage}
		& Surface \\
        \hline
		Brillouin Scattering & Distributed & \makecell{Optical Fiber\\Sensing Textile\cite{3214biondi2021optical,3215biondi2023smart}} &
		\begin{minipage}[t]{0.3\columnwidth}
			\centering
			\raisebox{-.5\height}{\includegraphics[width=\linewidth]{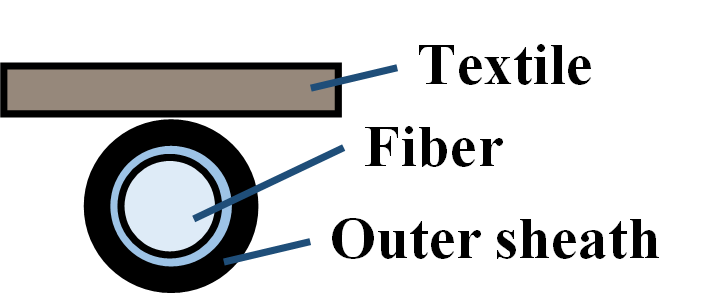}}
		\end{minipage}
		& Surface \\
		\hline
		Rayleigh Scattering & Distributed & \makecell{Embossed Surface\\Optical Cable\cite{3216alj2021environmental}} &
		\begin{minipage}[t]{0.5\columnwidth}
			\centering
			\raisebox{-.5\height}{\includegraphics[width=\linewidth]{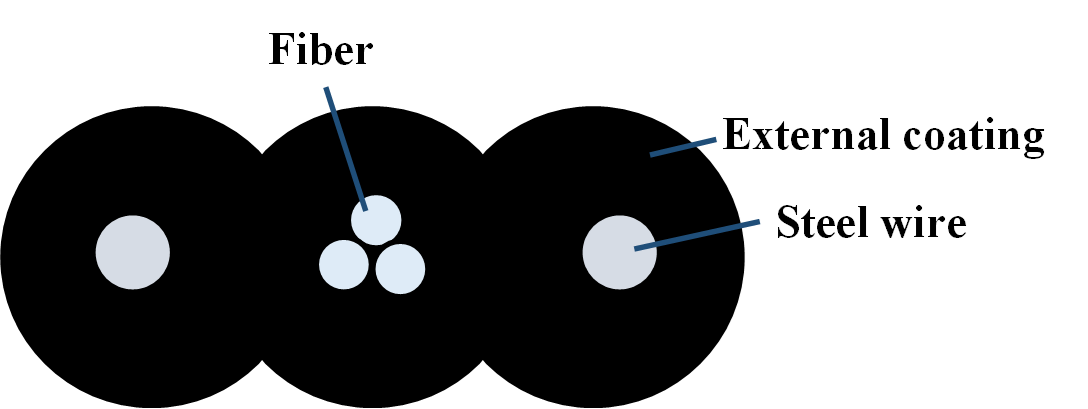}}
		\end{minipage}
		& Surface Embedded \\
		\hline
		Rayleigh Scattering & Distributed & \makecell{Temperature-Strain\\Composite Cable\cite{3210falcetelli2020strain,3211zhang2020situ}} &
		\begin{minipage}[t]{0.5\columnwidth}
			\centering
			\raisebox{-.5\height}{\includegraphics[width=\linewidth]{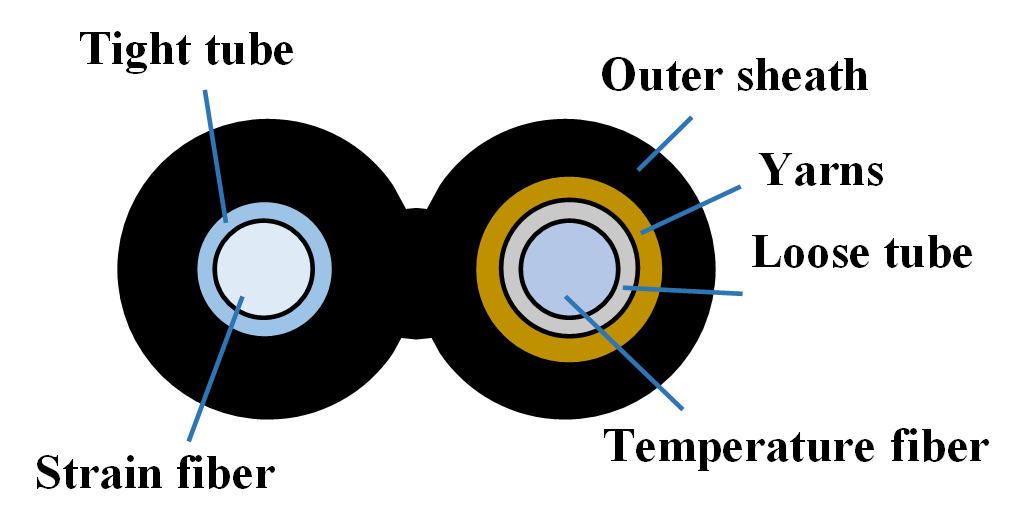}}
		\end{minipage}
		& Surface Embedded \\
		\hline
		FBG Array & Distributed & Long-Gauge Cable\cite{3208yue2024research,3209nan2024experimental} &
		\begin{minipage}[t]{0.5\columnwidth}
			\centering
			\raisebox{-.5\height}{\includegraphics[width=\linewidth]{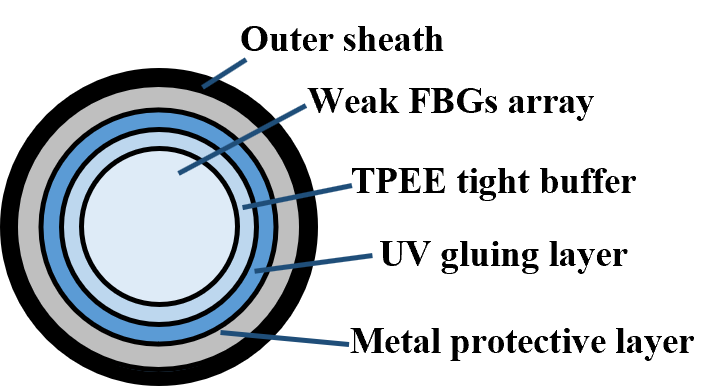}}
		\end{minipage}
		& Surface Embedded \\
		\hline
		FBG Array & Distributed & \makecell{High Mechanical Strength\\Long-Gauge Cable\cite{3218nan2025study}} &
		\begin{minipage}[t]{0.5\columnwidth}
			\centering
			\raisebox{-.5\height}{\includegraphics[width=\linewidth]{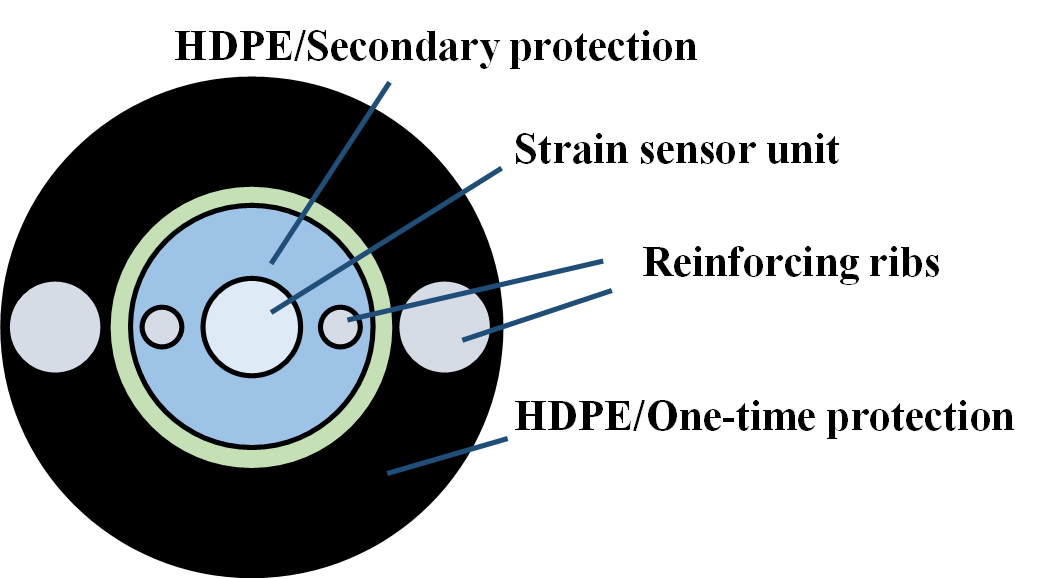}}
		\end{minipage}
		& Surface Embedded \\
		\hline
	\end{tabular}
\end{table*}

\subsubsection{Vibration/Acoustic Sensing Optical Cables}\label{2.2.3}
Distributed Fiber Optic Acoustic Sensing (DAS) technology, capable of fully distributed online monitoring, has become a crucial method for the structural health monitoring of large-scale transportation infrastructure. It enables the detection of minute structural anomalies across extensive areas. Through phase demodulation, it achieves a sensitivity three orders of magnitude higher than that of wavelength demodulation. The primary technical approaches include distributed sensing based on scattering effects (DAS) and sensing based on dual-beam interferometric arrays utilizing fiber grating arrays. These technologies facilitate the localization and identification of structural anomalies by analyzing the time-space characteristics of distributed acoustic wave signals under fixed excitation frequencies\cite{3220song2021sensing,3221hamanaka2024estimation}. They have been successfully applied in the health monitoring of major transportation infrastructure such as bridges\cite{3222cheng2019dynamic,3223golovastikov2025optical}, roads\cite{3224zhang2025expressway,3225an2023traffic}, and rail tracks\cite{3226madan2025monitoring,3227jiang2021real}.

In distributed fiber optic acoustic sensing systems, the cable encapsulation structure must simultaneously meet the dual requirements of high acoustic wave coupling efficiency and high mechanical reliability. Among these, tight-buffered encapsulation technology has been proven to significantly improve acoustic wave coupling efficiency. Comparative tests by Shang et al.\cite{3228shang2019discussion} showed that tight-buffered fibers offer superior acoustic sensitivity compared to bare fibers, and Hofmann et al.\cite{3229hofmann2015analysis} found that the tight buffer layer can effectively conduct strain waves and pressure waves. Furthermore, Rao et al.\cite{3230xiao2022review} proposed that using fibers with a tight-buffered encapsulation featuring a coating with high Young's modulus, low bulk modulus, and appropriate thickness can further enhance sensitivity. Current vibration sensing cable encapsulation technologies primarily achieve an optimal balance between sensitivity and strength through three types of structural design solutions: (1) Gel-filled loose-tube structures ensure low-frequency sensitivity through special buffer design, while the gel medium can effectively buffer mechanical impacts\cite{3231adeniyi2025directional}. (2) Polymer tight-buffered outer jackets\cite{3232han2020distributed,3233freeland2017relative} not only enhance acoustic wave coupling efficiency but also improve bending and tensile resistance through their material properties, allowing for a reduction in internal armor and reinforcement structures. (3) Aramid-reinforced tight-buffered structures eliminate sensitivity degradation by abandoning metal components, while leveraging the high-strength characteristics of aramid fibers to ensure cable reliability under complex working conditions. These new encapsulation solutions provide critical technical support for building fiber optic sensing systems that integrate high-sensitivity detection performance with engineering practicality. 

In practical engineering applications, the encapsulation design of vibration/acoustic sensing cables requires differentiated selection based on the specific monitoring scenario's sensitivity requirements and mechanical strength demands. Conventional tight-buffered cable structures are suitable for scenarios with lower mechanical strength requirements but a need for high-sensitivity detection. For example, Madan\cite{3226madan2025monitoring} and Li\cite{3234li2024monitoring} successfully monitored structural anomalies such as bolt loosening and pipeline wire breakage by employing PVC tight-buffered and standard tight-buffered cables, respectively, directly adhered to rail surfaces or buried near pipelines. Enhanced encapsulation designs are applicable to complex environments with higher mechanical strength requirements, necessitating structural optimization to balance high sensitivity with tensile, compressive, and shear resistance. Han et al.\cite{3231adeniyi2025directional} proposed reducing cable diameter and weight to improve sensitivity for buried monitoring. Dai et al.\cite{3235dai2025dynamic} replaced metal sheathing with a heat-shrink tubing layer, effectively enhancing the acoustic response of perimeter security cables. Furthermore, our research group developed a composite-encapsulated fiber grating array vibration sensing cable based on a buffer tube and GFRP (Glass Fiber Reinforced Polymer) strength member\cite{3236nan2019novel}. This structure demonstrates excellent engineering applicability in complex environments. It has successfully enabled precise identification of events such as personnel intrusion in urban rail transit\cite{3236nan2019novel} and fastener loosening\cite{3237li2022looseness}. In highway buried applications, it has identified various road defect hazards, achieving early warning for subsurface defects and road disasters\cite{3238li2025site}. These comparative studies indicate that cable encapsulation designs for different application scenarios must strike a balance between structural adaptability and performance optimality. Only then can distributed acoustic sensing systems achieve their best monitoring effectiveness in real-world engineering projects.

The sensing characteristics and structures of vibration/acoustic sensing optical cables are shown in Table \ref{tab6}. Due to the inconsistency in acoustic sensitivity testing methods and parameters across various reports, the table employs a star rating method (number of \ding{72} symbols) to provide a semi-quantitative comparison of the acoustic sensitivity of different optical fibers and cables.

\begin{table*}[!t]
	\caption{The Sensing Characteristics and Structures of Vibration/Acoustic Sensing Cables.\label{tab6}}
	\centering
	\begin{tabular}{cclcc}
		\hline
		\textbf{Sensing Principle} & \textbf{Cable Type} & \textbf{Typical Structure} & \textbf{\makecell{Installation\\Method}} & \textbf{\makecell{Acoustic\\sensitivity}} \\
		\hline
		\multirow{2}*{\makecell{\\ \\ \\ Rayleigh Scattering}} & Tight-Buffered Cable\cite{3226madan2025monitoring,3232han2020distributed} & 
		\begin{minipage}[t]{0.5\columnwidth}
			\centering
			\raisebox{-.5\height}{\includegraphics[width=\linewidth]{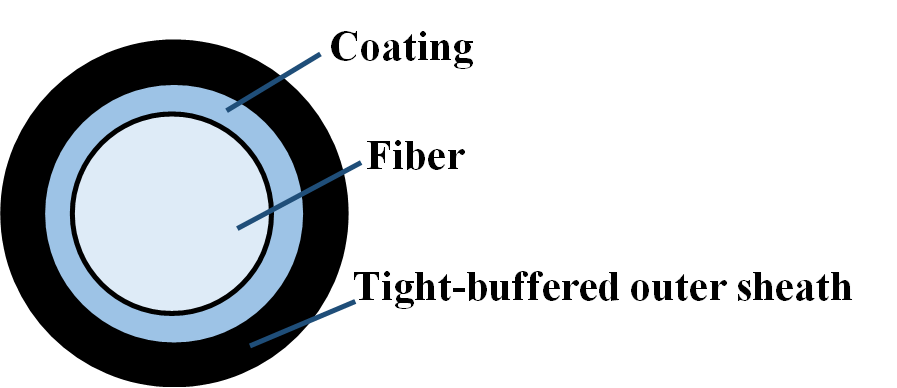}}
		\end{minipage}
		& Surface Embedded & \ding{72} \ding{72} \ding{72} \ding{72} \ding{73} \\
		\cline{2-5}
		  & Fibers Dispersed in Solid Polymer\cite{3231adeniyi2025directional} & 
		\begin{minipage}[t]{0.4\columnwidth}
			\centering
			\raisebox{-.5\height}{\includegraphics[width=\linewidth]{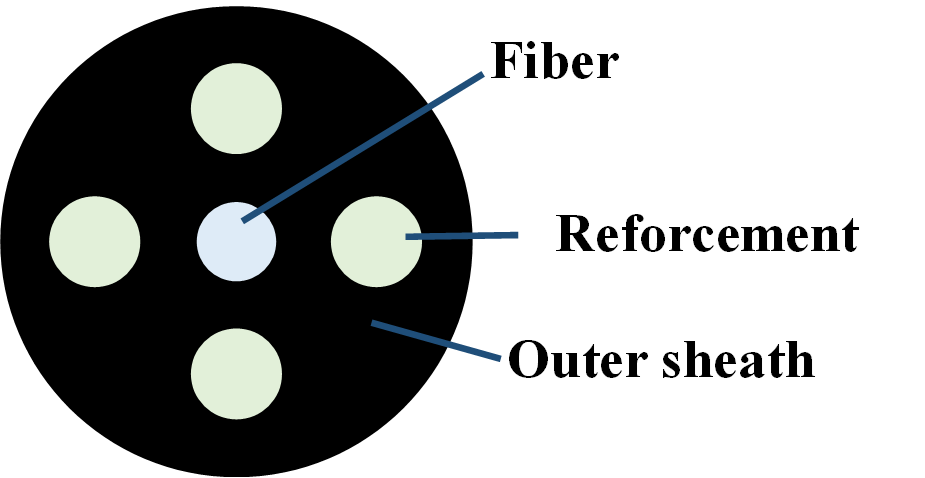}}
		\end{minipage}
		& Embedded & \ding{72} \ding{72} \ding{72} \ding{72} \ding{73} \\
		\hline
		\multirow{2}*{\makecell{\\ \\ \\ FBG Array}} & Enhanced GFRP Cable\cite{3236nan2019novel,3238li2025site} & 
		\begin{minipage}[t]{0.5\columnwidth}
			\centering
			\raisebox{-.5\height}{\includegraphics[width=\linewidth]{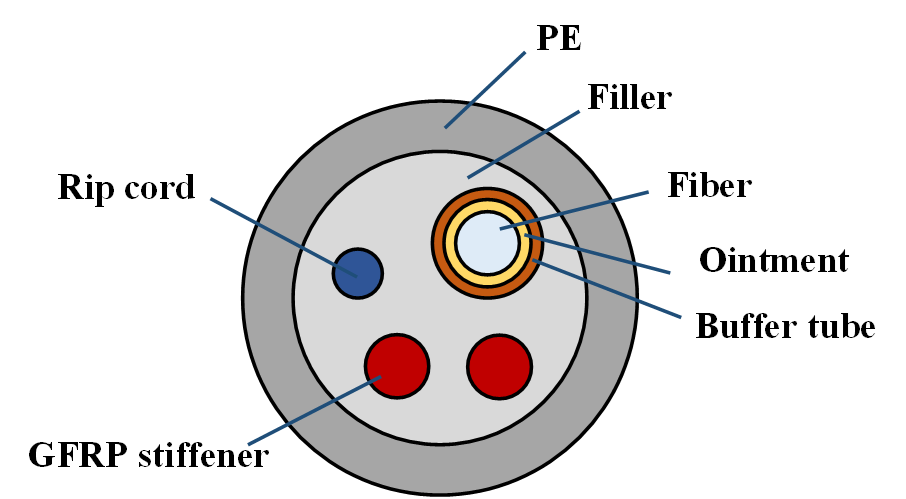}}
		\end{minipage}
		& Surface Embedded & \ding{72} \ding{72} \ding{72} \ding{72} \ding{73} \\
		\cline{2-5}
		& Metal Loose Tube Cable with Gel\cite{3225an2023traffic} & 
		\begin{minipage}[t]{0.5\columnwidth}
			\centering
			\raisebox{-.5\height}{\includegraphics[width=\linewidth]{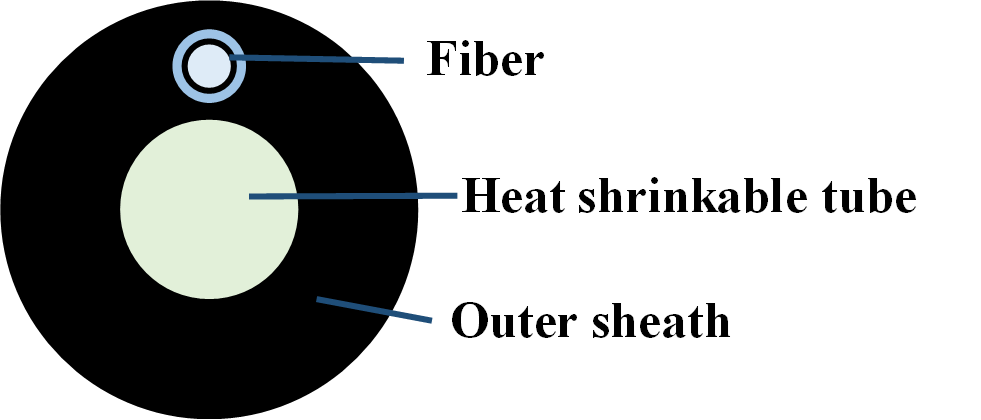}}
		\end{minipage}
		& Embedded & \ding{72} \ding{72} \ding{72} \ding{73} \ding{73} \\
		\hline
	\end{tabular}
\end{table*} 

\section{Recent Advances in OFS for Intelligent Transportation Infrastructures}\label{3}
The advances in OFS techniques have established extensive applications in civil infrastructures, particularly for the SHM demands where dense spatial coverage as well as tracking long term physical metrics including vibration, strain, and temperature, are crucial for safety-critical scenarios. Long-range transportation infrastructures such as railways, expressway, and bridges, favor OFS techniques due to its extraordinary sensing accuracy and robustness to complex environments. In this section, we review how OFS techniques address these capabilities in practical scenarios. We first investigate the conventional physical signal processing based OFS systems, focusing on transportation SHM deployments. Then we comprehensively discuss how ML methods are leveraged to process large-scale sensing data, enhance detection and classification accuracy that enables modern intelligent transportation infrastructures.

\subsection{Physics-Informed OFS for Transportation Infrastructures}\label{3.1}

Physics-informed OFS deployments in transportation infrastructures typically exploit the fundamentals of optical scattering and interference mechanisms including Rayleigh and Brillouin scattering, FBG-based reflection together with mechanical models of the target structure. In the following, we first review conventional OFS techniques for railway infrastructures, where long-range monitoring and safety-critical operations naturally corresponds to the capabilities of DOFS. Expressway and bridges applications are then discussed in subsequent subsections. 

\textbf{Smart Railway Monitoring}:
Railway systems are among the earliest transportation infrastructures to adopt OFS for distributed condition monitoring and traffic supervision. Existing studies in \cite{401sasi2020review,402minardo2013real,403wheeler2018measurement} summarize how FBG, Brillouin and Rayleigh scattering based DAS have been integrated into track as well as vehicles components to enable smart railway monitoring. 

\subsubsection{Train traffic and operation monitoring}\label{3.1.1}
Prior works focused on high fidelity local measurements of wheel–rail interactions using FBG arrays. Authors in \cite{404filograno2011real,405kouroussis2016railway} deployed FBG strain sensors on rail foot and sleeper components to achieve real-time monitoring of train passages, axle counting, speed estimation, and wheel flatness detection. Numerical results on commercial traffic networks demonstrated robust detection of trains running at up to 300 km/h with consistent axle counting and qualitative identification of defective wheels. While FBG arrays provide high resolution of local sensing, they are inherently constrained to discrete and sparse sensor locations. As a result, a significant portion of the rail line between FBGs remain uncovered, which yields coarse event localization for hazards. 

To address this limitation, authors in \cite{406kowarik2020fiber} exploited DAS on a $35$ km section of a high speed railway to monitor train trajectories continuously along a commercial optical telecom network, achieving a velocity standard deviation less than $5$ km/h for a train running at 160 km/h, and it can be reduced to $0.8$ km/h by further exploiting bogie-cluster signatures. Further studies in \cite{407fan2022high,408xie2023railway} push FBG-DAS railway monitoring towards large scale online defect detection. Numerical results validates that the proposed framework can precisely identify multiple track defects with a high-precision localization accuracy presented in maximum standard deviation within 0.314 meters. Their works address problems of DAS deployments regarding limited measurement range, low spatial resolution, and vulnerability to low SNR conditions by introducing a dynamic range extension phase unwrapping algorithm that improves received signal recovery over long distances.

\subsubsection{Track Substructure Condition Measurements and Intrusion Detection}\label{3.1.2}
For the underlying permanent track, OFS has been used to convert distributed strain and vibration measurements into rail bending indicators, sleeper reaction forces, and track support stiffness. Prior work in \cite{409velha2019monitoring} proposes a hybrid OFS framework for monitoring more than 50 kilometers of high speed railway infrastructure, in which a combination of FBG sensors and Raman-based distributed temperature sensing (RDTS) is applied. The proposed framework addresses the key problems of scaling high resolution local sensors to network-level coverage, and the lack of integrated temperature compensation and redundancy. And \cite{410parajuli2025rail} investigates DAS with track-mounted fibers for rail support condition monitoring. They monitor a 50 meters track segment instrumented with two types of optical cables bonded to the rail, while a locomotive traverses the section at speeds ranging from $16$ to $97$ km/h. This study tackles the main limitations of weak mechanical coupling in trench-laid DAS experiments by enforcing direct rail coupling and provides the first demonstrations of DAS for continuous track support monitoring.

Beyond railway routine traffic and condition monitoring, OFS has been increasingly used to detect safety-critical hazards such as broken rails, intrusions, and rockfalls. A series of recent DAS-based studies in \cite{411rahman2024remote,412han2024intelligent,413meng2020research} highlight the potential of distributed vibration sensing for predictive maintenance and anomaly detection along rail corridors. Particularly, authors in \cite{413meng2020research} address the high nuisance alarm rate (NAR) problem caused by unknown and time-varying vibration sources along the rail line. They implement a complete railway perimeter sensing chain, from the DAS front-end to the final intrusion decision, including data collection on an operational line, framing and denoising of $\varPhi$-OTDR traces, and extraction of hand-crafted vibration features that preserve the essential spatial-temporal characteristics of external events. Furthermore, they adopt an XGBoost-based classifier to distinguish multiple types of common intrusion events from benign background disturbances, obtaining an average recognition accuracy of $98.5\%$ and significantly reducing the NAR compared to the baselines.

\textbf{Expressway and Highway Monitoring}: Compared to railway monitoring, expressway and highway exhibit more heterogeneous traffic patterns, multilane layouts, and complex pavement structures. 

\emph{1)}\,\, \emph{Traffic and Vehicles Flow Monitoring:}
DAS-based traffic monitoring presented in \cite{414van2021next} proposes to calibrate propagation velocities and array responses, thus enhancing robustness in complex scenarios where simple thresholding or local time–frequency analysis \cite{415liu2018traffic} fails under overlapping signatures and strong environmental noise. Empirical studies with controlled vehicle runs show reliable detection and speed estimation over several hundred meters of highway, describing how array-level physical modeling can lift the limitations of previous single trace algorithms.

In comparison to DAS-based deployments, recent works in \cite{416jiang2025application,3238li2025site} introduce UWFBG sensing arrays into expressway pavements for axle and wheelbase detection.  They propose to place UWFBG vibration sensing cables in the asphalt base of expressway constructions, with the sensing cables laid vertical to the vehicles driving direction to form several instrumented cross sections. Experiments on multi-vehicle traffic flow scenarios demonstrate wheelbase errors below $1\%$ and reliable separation of densely spaced vehicles, addressing limitations of conventional piezoelectric quartz weigh-in-motion (WIM) sensors regarding electromagnetic interference, installation constraints, and maintenance costs. Building on this premise, reference \cite{418liu2024long} further extends UWFBG-based monitoring from localized cross sections to continuous long-distance traffic surveillance along the entire highway. In this work, the authors deploy multiple UWFBG arrays with more than $10000$ sensing points beneath the center of each lane of the Ezhou Huahu Airport highway, covering $7.08$ kilometers long and achieving lane-level spatial sampling at $5$ meters resolution. 

\emph{2)}\,\, \emph{Pavement and Subgrade Condition Assessment:} Beyond traffic flow sensing, OFS has been extensively applied to monitor pavement structural response and damage evolution. Reference \cite{419nosenzo2013continuous} presents the applications of FBG sensors for monitoring mining-induced strain in road pavements, where FBGs are embedded near the surface of an asphalt overlay to record strain histories during mining activities in the vicinity of the roadway. Extensive studies in \cite{420al2019weigh} develop a WIM system based on FBGs embedded in glass fiber-reinforced polymer (GFRP) bars within flexible pavements. The proposed system demonstrates superior alignments with reference sensors across a range of vehicle types and speed levels. Another line of recent FBG studies lies in full-domain monitoring systems that jointly track structural responses of highway bridges and the operational traffic states \cite{421yue2025research}. Real-world experiments in this work show the obtained vehicle speed errors on the order of a few percents over repeated round-trip tests and highlight practical challenges including temperature compensation strategies and time consumption.

While quasi-distributed FBGs provide high fidelity measurements at discrete points, their spatial sparsity motivates the use of distributed Brillouin sensing techniques to capture continuous strain distributions along pavements. Reference \cite{422bao2016strain} performs pulse pre-pump Brillouin optical time-domain analysis (PPP-BOTDA) to measure continuous strain distributions and detect cracks in thin unbonded concrete pavement overlays. Experiments at the MnROAD facility under repeated truck loading proves that PPP-BOTDA can capture crack initiation and propagation over the entire test section which effectively tackles the localization issues of isolated FBGs. This work presents a shift from point-based sensing to fully distributed pavement sensing and illustrates how BOTDR can be improved for overlay failure warning.

\textbf{Bridge and Viaduct Monitoring}: Bridge structures monitoring has emerged as another critical class of OFS application in transportation infrastructures. Recent SHM reviews for bridges, such as the integrated SHM survey by \cite{423he2022integrated}, position OFS alongside piezoelectric, GNSS and magnetostrictive sensors as the core sensing technologies in modern bridge monitoring frameworks. These systems aim to support long-term condition assessment, performance prediction, and early warning by integrating dense sensing, wireless communications and data analytics. 

\emph{1)}\,\, \emph{Long-term strain monitoring and integrated bridge SHM:} In general, Brillouin and Rayleigh-based DOFS can deliver kilometers-level coverage of strain and temperature on long bridges, but the key challenge arises with the trade-offs in spatial resolution and update rate. Practical studies in \cite{3213enckell2011evaluation} provide an example of Götaälv Bridge monitoring in Sweden, where 5 kilometers of Brillouin scattering based strain sensing fiber and 1 kilometers of temperature sensing fiber are deployed, to track strain, temperature and crack development on an aging steel girder bridge. The deployed system succeeds in identifying crack formation and abnormal structural behavior. 

To improve spatial resolution while retaining the long gauge lengths, references \cite{3208yue2024research,3209nan2024experimental} design weak FBG array strain optical cable specifically targets distributed strain measurement of bridges. Under mid-span loading up to $450$ kN, the grating array reproduces the expected strain distribution along the beam and aligns with conventional foil gauges. The coefficient of determination $R^2$ between grating-array readings and reference sensors exceeds $0.99$, while the error with respect to strain gauges decreases from about $28.3$\% at $100$ kN to $-1.9$\% at $450$ kN on the bottom plate, and stays fixed within roughly $\pm 6.5$\% compared to strong FBGs across $50-450$ kN. It is also reported that by post-processing the distributed strain via a conjugate beam method, deflection profiles can be reconstructed with satisfactory consistency. The key advantage over previous Brillouin DOFS systems is the denser quasi-distributed sampling along a single element.

\emph{2)}\,\,\emph{Dynamic Response and Train–Bridge Coupling:} For railway bridges and viaducts, dynamic response under moving trains and extreme conditions, e.g., earthquakes in seismically active areas, is the major safety concern. Authors in \cite{427cheng2024dynamic} reports a proof-of-concept deployment of an OFDR-based DOFS system on a masonry arch rail bridge in Gavirate. Upon train passages, the measured dynamic strains show characteristic negative peaks associated with locomotive and bogie loads, while the highest compressive peaks are typically in the range from $-160$ to $-170 \,\mu\varepsilon$. In comparison to long-gauge Brillouin or weak FBG systems, this work provides ultra-high spatial sensing resolution and it is well suited to masonry arches with localized cracking and complex stress paths. 

\begin{table*}[t]
	\caption{Physics-Informed OFS for Intelligent Transportation Infrastructures}
	\label{tab7}
	\centering
	\renewcommand\arraystretch{2} 
	\begin{tabular}{clll}
		\hline
		\textbf{Scenario} & \textbf{Scheme} & \textbf{Key Methods} & \textbf{Empirical Results}\\
		\hline
		
		\multirow{3}{*}{\textbf{Railway}}
		& FBG array
		& \makecell[l]{Wheel-rail interaction modeling and calibration, peak \\ thresholding and tracking~\cite{404filograno2011real,405kouroussis2016railway}}
		& \makecell[l]{Robust commercial lines up \\ to $300$ km/h} \\
		
		& DAS ($\varphi$-OTDR)
		& \makecell[l]{Iterative thermal-drift removal, trajectory inference \\ from spatio-temporal signatures, bidirectional elastic\\-wave tracking~\cite{406kowarik2020fiber,407fan2022high,408xie2023railway,411rahman2024remote,412han2024intelligent,413meng2020research}}
		& \makecell[l]{$35$ km railway line, velocity \\ std $<5$ km/h  at $160$ km/h}\\
		
		& Hybrid FBG + RDTS
		& \makecell[l]{Quasi-distributed FBG at key components + continuous \\ temperature profiling via RDTS for compensation and\\ redundancy~\cite{409velha2019monitoring}}
		& \makecell[l]{$>50$ km high speed \\ line monitoring} \\
		
		\hline
		
		\multirow{4}{*}{\textbf{Expressway}}
		& DAS
		& \makecell[l]{Beamforming processing to improve SNR and separate \\ multi-lane vehicles~\cite{414van2021next}.}
		& \makecell[l]{Reliable detection and speed \\ estimation over $100$ m} \\
		
		& UWFBG array
		& \makecell[l]{Interferometric UWFBG cells interrogated by OTDR, \\ trajectory reconstruction from spatio-temporal vibration \\ waterfall maps  \cite{416jiang2025application,3238li2025site,418liu2024long,419nosenzo2013continuous,420al2019weigh}}
		& \makecell[l]{Errors $<1\%$, $7.08$ km cover\\-age, lane-level sampling at $5$ \\ meters resolution}\\
		
		& PPP-BOTDA
		& \makecell[l]{Centimeter-level distributed Brillouin sensing \cite{422bao2016strain}}
		& \makecell[l]{Captures crack propagation \\ with cm resolution}\\
		
		\hline
		
		\multirow{3}{*}{\textbf{Bridge}}
		& Brillouin DOFS
		& \makecell[l]{Kilometer-scale Brillouin strain and temperature \\ monitoring \cite{3213enckell2011evaluation}}
		& \makecell[l]{mm level crack detection}\\
		
		& weak FBG cable
		& \makecell[l]{weak FBG for dense quasi-distributed strain, (beam \\ method)~\cite{3208yue2024research,3209nan2024experimental}.}
		& \makecell[l]{$R^2 > 0.99$ against foil gauges, \\ local strain sensitivity across \\ $1.21\sim1.43$ pm/$\mu\varepsilon$} \\
		
		& OFDR DOFS
		& \makecell[l]{High-resolution OFDR-based DOFS \cite{427cheng2024dynamic}}
		& \makecell[l]{Spatial resolution $2.56$ mm, \\ sampling up to $250$ Hz, peak \\ compressive strains $-170\sim$\\$-160 \mu\varepsilon$}\\
		
		\hline
	\end{tabular}
\end{table*}

To investigate coupled train–bridge dynamics under earthquakes, reference \cite{428xiang2024quasi} develops a quasi-distributed FBG sensing scheme for a $1/10$-scale high-speed train–bridge model. The proposed model presents a simple-supported girder bridge with the simulated layers of China Railway Track System type II (CRTS II) ballastless track slabs, mounted on a table shaking system.  The train model traverses the bridge at $9$ m/s and $13$ m/s, respectively, which corresponds to $100$ km/h and $150$ km/h speed in full scale, while sinusoidal horizontal and combined horizontal–vertical seismic excitations are applied with spectrum of $8–15$ Hz and peak ground accelerations (PGA) of $0.1$ g and $0.2$ g.

In conclusion, railway, expressway and bridge applications studies in Table \ref{tab7} show that physics-informed OFS deliver kilometers-scale coverage, high fidelity strain and vibration fields, and physically interpretable indicators such as axle loads, track stiffness, and bridge deformation. 

\subsection{Machine Learning-Enabled OFS for Intelligent Transportation Infrastructures}\label{3.2}

While physics-informed OFS has demonstrated strong capability in modern transportation infrastructures monitoring, practical deployments are still constrained by extremely large scale sensing data, heterogeneous traffic patterns, and complex propagation environments. Conventional approaches often rely on handcrafted features, simplified mechanical models, and thresholding designs which may incur scalability issues, unreliable performance under non-stationary conditions, and high calibration costs. 

ML-aided schemes introduce a fundamentally new paradigm by learning hierarchical representations directly from massive OFS data, enabling sensing systems to evolve from physics logit interpretation to data-driven infrastructure perception. Rather than treating OFS merely as a measurement tool, ML-enabled OFS frameworks emerges as intelligent perception systems that automatically extract latent structures from raw spatio-temporal measurements, discriminate complex events, and support scalable monitoring. In the following, we review how ML techniques are integrated with OFS to promote system-level capabilities in railway, expressway, and bridge infrastructures.

\textbf{Data-Driven OFS for Smart Railway}:
Compared to physics-informed railway monitoring systems that  focus on signal detection and parameter estimation, learning-based OFS frameworks target on task-oriented perception, including intelligent intrusion recognition, defect diagnosis, and operational state interpretation. These developments depict a shift from ``whether an event occurs" to ``what the event is and how it evolves".

\emph{1)}\,\, \emph{Intelligent Intrusion Recognition:}
Prior works formulate DAS measurements as large scale spatio-temporal data and adopt deep neural networks (DNN) to learn discriminative representations directly from raw traces or time–frequency maps. Reference \cite{411rahman2024remote} proposes a hybrid convolutional–recurrent architecture which actively learns spectral–temporal features from DAS signals for railway condition monitoring. By integrating CNN feature extraction with LSTM temporal sequence modeling, the system achieves robust recognition of railway events, reporting recognition accuracy exceeding $97\%$ while maintaining stable performance under strong background noise.

Similarly, a series of prior works \cite{429yurekli2025real,430chiang2024distributed,431shiloh2019efficient} adopt CNN trained on DAS time–frequency representations to classify railway intrusions and operational events. Empirical results on real-world railway scenarios present classification accuracies above $95\%$ across multiple event categories, with substantially improved separability compared to handcrafted feature baselines. These studies collectively demonstrate that ML techniques transform DAS-based railway monitoring systems from passive vibration detectors into perception modules capable of interpreting semantic events.

\emph{2)}\,\, \emph{ML-Aided Defect Diagnosis and Track Condition Assessment:}
Beyond perimeter monitoring, ML-aided OFS systems also address the problems of defect-oriented railway diagnostics, where explicit mechanical modeling becomes unreliable due to coupling uncertainty, heterogeneous track structures, and overlapping vibration signatures. Existing studies explore the integration of deep learning and DAS measurements to identify loose fastener states along operational tracks \cite{412han2024intelligent,432dong2025heterogeneous}. Empirical results indicate that the deep learning-based systems can achieve detection accuracies above $96\%$, which significantly outperform conventional energy-based and handcrafted features approaches. Moreover, the proposed frameworks enable continuous kilometer-scale monitoring without requiring component-level physical modeling, showing how ML enables OFS systems to move from indirect vibration sensing toward direct infrastructure condition perception.

In a related line of works on DAS-based defect recognition further show that deep learning-based systems can capture subtle defect induced patterns that are difficult to separate through classical filtering and propagation fitting \cite{433wu2025high,434wang2023novel,435wang2023wheel}. Multi-scale convolutional networks and attention mechanism are applied to DAS measurements to identify defect related signatures, achieving accuracies above $96\%$ under controlled and semi-operational conditions. Such results indicate that ML substantially expands the functional envelope of railway OFS systems from anomaly indication to reliable defect diagnosis.

\textbf{Learning Based OFS for Smart Expressway Monitoring:}
Expressway infrastructures monitoring faces distinct challenges compared to railway scenarios, since they involve multi-lane traffic flows, diverse vehicle types, and continuous spatio-temporal patterns spanning long distance. While physics-informed OFS approaches can extract coarse traffic signatures, they often struggle with robustness and scalability in real-world roadway characterized by overlapping vehicle responses, strong environmental noise, and non-stationary traffic dynamics. Learning based OFS frameworks are therefore introduced to directly extract semantic representations for providing robust traffic event separation and unified perception across extended highway.

\emph{1)}\,\, \emph{ML-Enabled Traffic Flow Perception:}
Existing studies fall into the categories of transforming DAS-based highway monitoring from simple vehicle detection to structured traffic perception, including vehicles recognition, flow estimation, and speed analysis under ultra-dense traffic conditions. References \cite{436van2022deep,437khacef2025precision} propose to embed DNN into DAS-based highway monitoring pipelines not only as post-processing classifiers, but also as core perception components. A deep deconvolution framework is introduced to learn the inverse mapping from raw DAS measurements to high fidelity spatio-temporal traffic representations. By doing so, they improve the separability of closely spaced vehicles and overlapping responses under dense traffic flow.  Real-world experiments demonstrate reliable multi-vehicle discrimination with false detection rate around $5.22\%$. These studies collectively establish a representative ML-enabled highway OFS paradigm, where learning-based models improve effective sensing resolution and stabilize long distance monitoring.

Beyond perception-oriented traffic analysis, ML methods have also been adopted to reduce the latency of expressway OFS systems, especially for lane-level identification where vibration signatures are strongly coupled across lanes and the sensing environment is highly non-stationary. Authors in \cite{438liu2023vehicle} develop a learning-based monitoring pipeline in which UWFBG arrays are deployed underground in each lane to capture vibrations, and a DBSCAN-based procedure is used to extract three types of vibration samples, i.e., single vehicle signals, accompanying signals, and laterally adjacent lane interference signals, for model training. A teacher–student architecture is then designed, where the teacher model combines ResNet feature extraction with LSTM temporal modeling. While the student model is simplified to a lightweight single layer LSTM and trained via knowledge distillation to meet latency constraints. Experimental results show that the distilled student model achieves an average vehicle identification rate of $95\%$ within acceptable latency.

\emph{2)}\,\, \emph{Learning-Based Feature Enhancement and Robust Highway Sensing:}

Learning-based models can also be used to improve the quality and separability of DAS representations, thereby strengthening downstream monitoring performance. In \cite{439yuan2023spatial}, a deep deconvolutional U-Net architecture is proposed to reconstruct high fidelity spatio-temporal representations from DAS measurements. Empirical studies show that the proposed architecture improves the separability of closely traveling vehicles and benefits the extraction of vehicle related patterns, e.g., axle-induced signatures, providing more reliable identification of traffic behaviors under congested highway conditions.

For low SINR scenarios where DAS measurements are severely contaminated by environmental disturbances and heterogeneous vibration sources, a learning-enhanced DAS traffic monitoring framework is developed in \cite{440wang2025enhancing} which explicitly targeting noise dominated highway scenarios. In this work, roadside DAS data are transformed into spatio-temporal representations and processed via a deep learning detection and classification network optimized over noisy vibration fields. Experiments demonstrate that the proposed method achieves traffic event classification accuracy of around $92\%$ under strong environmental interference, substantially outperforming conventional baselines. Moreover, the learning-based framework maintains stable detection performance across varying noise conditions and traffic densities, showing that deep learning can effectively suppress background disturbances and disentangle overlapping traffic signatures.

\textbf{Learning-Based OFS for Smart Bridge and Viaduct Monitoring:}
Bridge and viaduct infrastructures monitoring mainly focus on long term structural integrity, damage evolution, and safety assessment. Although physics-informed OFS systems have demonstrated strong capability in distributed strain and vibration monitoring, practical deployments still face fundamental challenges, such as heterogeneous structural responses, strong environmental shift, and the difficulty of extracting damage-sensitive information from massive spatio-temporal measurements. 

Existing studies focus on transforming distributed strain and vibration measurements into interpretable structural condition indicators. Reference \cite{441lu2025deep} proposes a physics-constrained learning model (PC-FiberNet) to interpret DOFS strain distributions under multi-crack scenarios. Rather than relying on handcrafted peak-picking rules, the framework is designed to directly infer crack locations and crack widths even when strain peaks overlap and when cable structure interface behaviors introduce strong nonlinearities which are common in real civil infrastructures. This study illustrates how learning-based interpretation can turn DOFS measurements into actionable damage descriptors, thereby enabling more automated and scalable crack assessment for bridge structures.

\begin{table*}[t]
	\caption{Machine Learning-Enabled OFS for Intelligent Transportation Infrastructures}
	\label{tab8}
	\centering
	\renewcommand\arraystretch{2}
	\begin{tabular}{clll}
		\hline
		\textbf{Scenario} & \textbf{Key Methods} & \textbf{Empirical Results} & \textbf{System-Level Gains} \\
		\hline
		
		\multirow{2}{*}{\textbf{Railway}}
		& \makecell[l]{CNN--LSTM hybrid networks and deep time--frequency \\ representation learning for DAS  perception~\cite{411rahman2024remote,429yurekli2025real,430chiang2024distributed,431shiloh2019efficient}}
		& \makecell[l]{Recognition accuracy\\ exceeding $95\%$}
		& \makecell[l]{From vibration detection\\ to semantic railway\\ event perception} \\
		
		& \makecell[l]{Integrating multi-scale CNNs and attention mechanisms for \\ defect-oriented representation learning~\cite{412han2024intelligent,432dong2025heterogeneous,433wu2025high,434wang2023novel,435wang2023wheel}}
		& \makecell[l]{Defect detection\\ accuracy above $96\%$}
		& \makecell[l]{From anomaly indication\\ to reliable defect diagnosis} \\
		
		\hline
		
		\multirow{3}{*}{\textbf{Expressway}}
		& \makecell[l]{Deep deconvolution and perception networks embedded \\ into DAS pipelines for structured traffic sensing~\cite{436van2022deep,437khacef2025precision}}
		& \makecell[l]{False detection rate\\ around $5.22\%$}
		& \makecell[l]{From vehicle detection\\ to structured traffic\\ perception} \\
		
		& \makecell[l]{Teacher--student architectures with knowledge distillation \\ for real time lane-level monitoring~\cite{438liu2023vehicle}}
		& \makecell[l]{Vehicle identification\\ rate around $95\%$ with \\ acceptable latency}
		& \makecell[l]{Real time and scalable\\ expressway perception} \\
		
		& \makecell[l]{Learning-based representation enhancement and robust deep\\ detection for DAS traffic sensing~\cite{439yuan2023spatial,440wang2025enhancing}}
		& \makecell[l]{Classification accuracy\\ around $$92\%$$ under\\ strong interference}
		& \makecell[l]{Robust to environmental \\ noise} \\
		
		\hline
		
		\multirow{2}{*}{\textbf{Bridge}}
		& \makecell[l]{Physics-constrained learning and deep detection models for \\ DOFS-based crack damage perception~\cite{441lu2025deep,110liu2023intelligent}}
		& \makecell[l]{Crack detection\\ mAP $=0.968$,\\ processing time less \\ than $0.05$ seconds for \\ $10{,}000$ points}
		& \makecell[l]{From distributed strain\\ sensing to actionable\\ damage descriptors} \\
		
		& \makecell[l]{Learning-based feature enhancement, representation learn\\-ing,  and anomaly-oriented perception for long term OFS \\ monitoring~\cite{443shao2025artificial}}
		& \makecell[l]{Stable representations\\ under long term and\\ multi-condition opera\\-tions}
		& \makecell[l]{Long term intelligent\\ structural perception} \\
		
		\hline
	\end{tabular}
\end{table*}

As a complementary evidence, authors in \cite{110liu2023intelligent} show that DFOS strain distributions can be transformed into strain-contour representations and processed with a modified deep YOLO detection model to identify and localize multiple cracks in real time. Experiments report a mean average precision of $0.968$ for crack detection, and the processing time for one DOFS dataset containing $10,000$ measurement points is less than $0.05$ seconds, highlighting both high accuracy and real-time feasibility. These results demonstrate that deep learning can serve as an effective perception module for crack monitoring with DOFS.

From a system-level perspective, the comprehensive review \cite{443shao2025artificial} highlights that learning-based feature enhancement, representation learning, and anomaly detection have emerged as essential components of modern OFS infrastructures. The review summarizes how ML techniques are increasingly adopted to suppress environmental interference, stabilize distributed sensing features, and construct latent structural representations that remain consistent across long term operations. These capabilities are identified as critical enablers for transitioning OFS deployments from experimental demonstrations toward task-oriented intelligent bridge monitoring systems.

In conclusion, ML-aided OFS constructs a unified intelligent sensing paradigm across railway, expressway, and bridge infrastructures, in which DOFS networks evolve from conventional signal acquisition systems into data-driven perception platforms capable of semantic interpretation, fine-grained condition diagnosis, and long term monitoring, thereby providing a basis for scalable, autonomous, and reliable intelligent transportation monitoring.

\section{Conclusion}\label{4}
DOFS has emerged as a critical tool for large scale infrastructures SHM, owing to its unique capabilities for long distance, wide coverage, and multi-modal sensing. This paper has systematically investigated the key advances in this field, encompassing main aspects such as distributed interrogation methods, the fabrication of specialty sensing fibers, the design of multi-modal cable packaging, and intelligent signal processing algorithms. By analyzing application cases in typical linear infrastructures including bridges, tunnels, and rail transit systems, the reliability and adaptability of DOFS in practical deployments have been validated. Empirical results demonstrate that DOFS maintains real time structural data with high precision, building a solid foundation for the intelligent operation and maintenance of modern infrastructures.

However, several challenges remain unsolved in transitioning DOFS towards scalable and intelligent deployments. First of all, existing OFS systems exhibit high sensitivity to environmental noise, making it hard to build an accurate mapping between collected raw signals and structural anomalies. Reducing Type-I and Type-II errors without standardized databases and reliable machine learning models remains a critical hurdle. On the other hand, the massive volume of monitoring data raises rigorous demands on hardware, where achieving miniaturized integration of interrogation equipment and improving algorithmic efficiency are significant bottlenecks for real time pattern analysis. Meanwhile, due to the diversity of sensing principles and the complexity of application scenarios, standardized solutions tailored to specific application contexts remain to be addressed.

Looking forward, we expect future works to focus on the following aspects: 1) enhancing the sustainability and reliability of sensing cables under extreme environments; 2) advancing hardware integration technologies for interrogation systems; 3) developing reliable and efficient structural health diagnostic methods with the aid of statistical learning \cite{444zhang2024bayesian,445balcan2023reliable,446jin2025invariant} and multi-fidelity learning techniques \cite{447zhang2025multi,448wu2023disentangled}; 4) building a standardized framework for object-oriented OFS applications. Technical breakthroughs in these areas will further catalyze the deep integration and industrialization of DOFS in the field of smart infrastructure monitoring. We hope that the insights provided in this review can shed the light for subsequent research and practical deployments in this subject.



\bibliographystyle{IEEEtran} 
\bibliography{ref2.bib}

\vfill

\end{document}